\documentclass[9pt,nocopyrightspace]{sigplanconf}

\usepackage[utf8]{inputenc}

\usepackage{epsfig}
\usepackage{epstopdf}
\usepackage[labelfont=bf]{caption}
\DeclareCaptionFormat{myformat1}{\hrulefill\newline#1#2#3}
\usepackage{subcaption}
\DeclareCaptionFormat{myformat2}{#1#2#3}
\usepackage{amsmath}
\usepackage{balance}
\usepackage{times}

\usepackage{xspace}

\usepackage{booktabs}
\newcommand{\tabhead}[1]{ \textbf{#1} }
\usepackage[table]{xcolor}
\definecolor{oddcolor}{HTML}{FFFFFF}
\definecolor{evencolor}{HTML}{CCD0FF}
\definecolor{rulesbgcolor}{HTML}{F8FFCC}
\definecolor{ruleslncolor}{HTML}{555555}

\definecolor{refColor}{HTML}{A63603}
\definecolor{citeColor}{HTML}{006D2C}
\definecolor{urlColor}{HTML}{08519C}

\usepackage[pagebackref=false,breaklinks=true,letterpaper=true,colorlinks,linkcolor=refColor,citecolor=citeColor,urlcolor=urlColor,bookmarks=false]{hyperref}

\usepackage[backgroundcolor=rulesbgcolor,linecolor=ruleslncolor,innerleftmargin=0pt,innerbottommargin=0pt, innertopmargin=-3pt, skipbelow=0pt,skipabove=0pt]{mdframed}

\usepackage{tikz}
\usepgflibrary{arrows}

\usepackage{bm}

\usepackage[T1]{fontenc}
\usepackage[scaled=0.9]{DejaVuSansMono}

\usepackage{syntax}

\usepackage{listings}
\usepackage{soul}

\usepackage{enumitem}

\begin{document}

\setlength{\pdfpageheight}{\paperheight}
\setlength{\pdfpagewidth}{\paperwidth}

\conferenceinfo{CONF 'yy}{Month d--d, 20yy, City, ST, Country} 
\copyrightyear{20yy} 
\copyrightdata{978-1-nnnn-nnnn-n/yy/mm} 
\doi{nnnnnnn.nnnnnnn}

\title{Patterns and Rewrite Rules for Systematic Code Generation}
\subtitle{From High-Level Functional Patterns to High-Performance OpenCL Code\vspace{1em}}

\authorinfo{Michel Steuwer\vspace{.5em}}
           {The University of Edinburgh\\ University of Muenster\vspace{.5em}}
           {michel.steuwer@ed.ac.uk}
\authorinfo{Christian Fensch\vspace{.5em}}
           {Heriot-Watt University\vspace{.5em}}
           {c.fensch@ed.ac.uk}            
\authorinfo{Christophe Dubach\vspace{.5em}}
           {The University of Edinburgh\vspace{.5em}}
           {christophe.dubach@ed.ac.uk}

\maketitle


\newcommand{\ie}{\emph{i.e.},\xspace}
\newcommand{\eg}{\emph{e.g.},\xspace}
\newcommand{\aka}{\emph{a.k.a.},\xspace}
\newcommand{\etal}{et al.\xspace}
\newcommand{\TODO}[1]{\textbf{\hl{TODO: #1}}}
\newcommand{\pat}[1]{\textit{#1}}
\newcommand{\bench}[1]{\textit{#1}}

\newcommand{\code}[1]{{\small\texttt{#1}}}

\thispagestyle{empty}

%
%
\begin{abstract}

Computing systems have become increasingly complex with the emergence of heterogeneous hardware combining multicore CPUs and GPUs. 
These parallel systems exhibit tremendous computational power at the cost of increased programming effort.
This results in a tension between achieving performance and code portability.
Code is either tuned using device-specific optimizations to achieve maximum performance or is written in a high-level language to achieve portability at the expense of performance.

We propose a novel approach that offers high-level programming, code portability and high-performance.
It is based on algorithmic pattern composition coupled with a powerful, yet simple, set of rewrite rules.
This enables systematic transformation and optimization of a high-level program into a low-level hardware specific representation which leads to high performance code.

We test our design in practice by describing a subset of the OpenCL programming model with low-level patterns and by implementing a compiler which generates high performance OpenCL code.
Our experiments show that we can systematically derive high-performance device-specific implementations from simple high-level algorithmic expressions.
The performance of the generated OpenCL code is on par with highly tuned implementations for multicore CPUs and GPUs written by experts.

\end{abstract}


\keywords
Algorithmic Patterns, Rewrite Rules, Performance Portability, GPU, OpenCL, Code Generation

\bigskip
\section{Introduction}

Computing systems have become extremely complex and diversified implementing different forms of parallelism and memory hierarchies.
Modern multicore CPUs and GPUs (Graphics Processing Units) are often used for general purpose computations.
The drawback of such systems is the extreme difficulty of programming and extracting performance, requiring a deep understanding of the hardware.
Software written and tuned for today's systems needs to be adapted frequently to keep pace with ever changing hardware.

Over the years, a wide range of languages, language extensions and frameworks have emerged for programming GPUs and other massively parallel devices. 
The two most common languages are CUDA and OpenCL, both directly exposing low-level hardware features.
Directive based approaches such as Open\-ACC~\cite{reyes12openaccgpu} and OpenMP~\cite{lee09openmp}, extensions to existing programming languages such as Cilk~\cite{blumofe95cilk} or libraries like Intel TBB~\cite{reinders07inteltbb} have been proposed to reduce the complexity of developing code for multicore CPUs and GPUs.
While these latter approaches simplify the development of applications, they all lead to an explosion of specialized implementations where the same algorithmic concept is tuned differently for each device.
As a result, \emph{performance portability} remains elusive; code optimized for one device might only achieve a fraction of the performance on a different device.

Several high-level programming models have been proposed to address this issue.
Petabricks~\cite{phothilimthana13portable} allows the programmer to express different algorithm implementations and automatically picks the best one using auto-tuning.
Higher-level dataflow programming language such as StreamIt~\cite{hormati11sponge} or LiquidMetal~\cite{dubach12compiling} have been designed with a similar goal in mind.
Both languages use dedicated backend compiler for different hardware targets such as GPUs.
Nvidia's NOVA~\cite{collins14nova} language takes a more functional programming approach where algorithmic patterns such as \pat{map} or \pat{reduce} are expressed as primitives recognized by the backend compiler.
While definitively a step in the right direction, all these approaches rely on ad-hoc techniques such as hard-coded device-specific implementations or heuristics.
When hardware changes occur, the backend compiler has to be re-tuned or re-engineered.

The root of the problem lies in a gap in the system stack between high-level algorithmic concepts on the one hand and low-level hardware paradigms on the other hand.
In this work we propose to bridge this gap by defining a set of rewrite rules which systematically translates high-level algorithmic concepts into low-level hardware paradigms, both expressed as functional patterns.
The rewrite rules are used to systematically derive semantically equivalent low-level expressions from high-level algorithm expressions written by the programmer.
Once derived, we can automatically generate high performance code based on these expressions.
Our approach is similar in spirit to Spiral~\cite{pueschel05spiral}, but relies on fine grain hardware patterns representing CPU and GPU hardware features. 
As a result, in our approach code generation becomes very simple since all optimization decisions are handled during the automatic rewriting process and no complex analysis is performed.

The power of our approach lies in the rewrite rules, written once by an expert system designer.
These rules encode the different algorithmic choices and low-level hardware specific optimizations.
The rewrite rules play the dual role of enabling the composition of algorithmic patterns and enabling the lowering of these patterns onto the low-level hardware paradigms.
This results in a clear separation of concerns between high-level algorithmic patterns and low-level hardware paradigms.
The rewrite rules define an implementation space that can be systematically searched to produce high performance code.
We believe these principles pave the way to fully automated portable high performance code generation.

\noindent
This paper demonstrates the practicality of our approach using OpenCL as our target hardware platform.
We compare our approach with highly-tuned linear algebra functions extracted from the state-of-the-art libraries and with larger benchmarks such as BlackScholes.
We express them as compositions of high-level algorithmic patterns which are systematically lowered to low-level OpenCL patterns from which OpenCL code is generated.
The performance of our generated code is competitive with highly-tuned BLAS linear algebra libraries such as Nvidia GPU CUBLAS, AMD GPU clBLAS and Intel MKL on the CPU.

Our paper makes the following key contributions:

\begin{itemize}[itemsep=2pt,parsep=1pt,topsep=2pt]
  \item design of \textbf{high-level algorithmic patterns} used by the programmer and \textbf{low-level OpenCL patterns} representing the OpenCL programming model;
  \item develop a powerful set of \textbf{rewrite rules} that systematically expresses algorithmic and optimization choices;
  \item achieve \textbf{performance portability} by systematically applying rewrite rules to derive device-specific implementations, leading to performance on par with the best hand-tuned versions.
\end{itemize}

\noindent
The paper is structured as follows.
Section~\ref{back} provides a motivation.
Sections~\ref{pattern} and~\ref{rules} present our patterns and rewrite rules.
Section~\ref{benchmarks} and~\ref{application} show our benchmarks and rules in action.
Our experimental setup and performance results are shown in Sections~\ref{sec:setup} and~\ref{sec:results}.
Finally, Section~\ref{related} discusses related work and Section~\ref{conc} concludes.


\section{Motivation}
\label{back}

\begin{figure}[t]
\centering
\includegraphics[width=\linewidth]{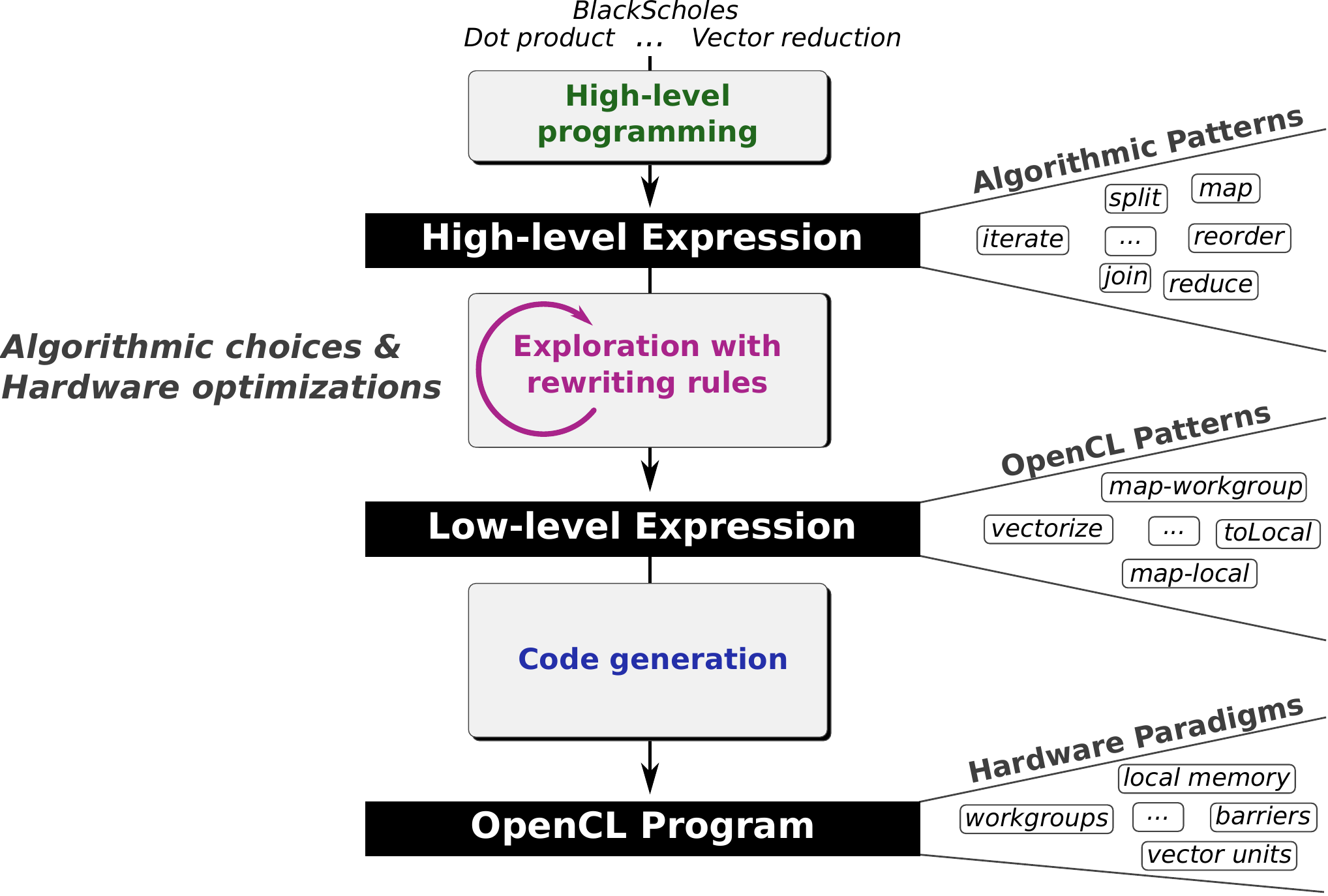}
\vspace{-19pt}
\caption{
Overview of our system.
The programmer expresses the problem with high-level algorithmic patterns.
These are systematically transformed into low-level OpenCL patterns using a rule rewriting system.
OpenCL code is generated by mapping the low-level patterns directly to the OpenCL programming model representing hardware paradigms.
\vspace{-1em}}
\label{fig:highlevel}
\end{figure}

The overview of our approach is presented in Figure~\ref{fig:highlevel}.
The programmer writes a \emph{high-level expression} composed of \emph{algorithmic patterns}.
Using a rewrite rule system, we systematically lower this high-level expression into a \emph{low-level expression} consisting of \emph{OpenCL patterns}.
In this rewrite stage algorithmic and optimization choices in the high-level expression can be explored.
The generated low-level expression is then fed into our code generator that emits an \emph{OpenCL program}.
This program is finally compiled to machine code by the vendor provided OpenCL compiler.


\begin{figure}[t]
\centering

\begin{subfigure}[b]{.85\linewidth}
\begin{lstlisting}[mathescape,numbers=left]
def mul3(x) = x * 3    // user-defined function
def vectorScal = map(mul3)       // map pattern
\end{lstlisting}
\caption{\textbf{High-level expression} written by the programmer.}
\label{fig:codeex:map}
\end{subfigure}

\vspace{-10pt}
\begin{minipage}{0.1\linewidth}
\vspace{0pt}
\centering
\begin{tikzpicture}[ultra thick]
 \draw [black,   -latex      ] (0,0.5) -- (0,0) node [] {};
\end{tikzpicture}
\end{minipage}
\begin{minipage}{0.25\linewidth}
\vspace{-5pt}
\centering
\textbf{rewrite rules}
\end{minipage}
\begin{minipage}{0.1\linewidth}
\vspace{0pt}
\centering
\begin{tikzpicture}[ultra thick]
 \draw [black,   -latex      ] (0,0.5) -- (0,0) node [] {};
\end{tikzpicture}
\end{minipage}

\vspace{0pt}
\begin{subfigure}[b]{\linewidth}
\centering
\begin{minipage}{.85\linewidth}
\begin{lstlisting}[mathescape,numbers=left]
def mul3(x) = x * 3
def vectorScal = join $\circ$ map-workgroup(
                  asScalar $\circ$ map-local(
                    vectorize-4(mul3)
                  ) $\circ$ asVector-4
                 ) $\circ$ split-1024
\end{lstlisting}
\end{minipage}
\caption{\textbf{Low-level expression} systematically derived using rewrite rules.}
\label{fig:codeex:impl}
\end{subfigure}

\vspace{-10pt}
\begin{minipage}{0.1\linewidth}
\vspace{0pt}
\centering
\begin{tikzpicture}[ultra thick]
 \draw [black,   -latex      ] (0,0.5) -- (0,0) node [] {};
\end{tikzpicture}
\end{minipage}
\begin{minipage}{0.26\linewidth}
\vspace{-5pt}
\centering
\textbf{code generator}
\end{minipage}
\begin{minipage}{0.1\linewidth}
\vspace{0pt}
\centering
\begin{tikzpicture}[ultra thick]
 \draw [black,   -latex      ] (0,0.5) -- (0,0) node [] {};
\end{tikzpicture}
\end{minipage}

\vspace{0pt}
\begin{subfigure}[b]{\linewidth}
\centering
\begin{minipage}{.85\textwidth}
\begin{lstlisting}[mathescape,numbers=left]
int4 mul3(int4 x) { return x * 3; }
kernel vectorScal(global int* in,out, int len){
 for (int i=get_group_id; i < len/1024;
      i+=get_num_groups) {
  global int* grp_in  = in+(i*1024);
  global int* grp_out = out+(i*1024);
  for (int j=get_local_id; j < 1024/4;
       j+=get_local_size) {
    global int4* in_vec4 =(int4*)grp_in+(j*4);
    global int4* out_vec4=(int4*)grp_out+(j*4);
    *out_vec4 = mul3(*in_vec4);      
} } }  
\end{lstlisting}
\end{minipage}
\caption{\textbf{OpenCL program} produced by our code generator.}
\label{fig:codeex:ocl}
\end{subfigure}
\vspace{-20pt}
\caption{
Pseudo-code representing vector scaling.
The user simply maps the \code{mul3} function over the elements of the input array~(\subref{fig:codeex:map}).
This high-level expression is systematically transformed into a low-level expression~(\subref{fig:codeex:impl}) using rewrite rules.
Finally, our code generator turns the low-level expression into an OpenCL program~(\subref{fig:codeex:ocl}).
}
  \label{fig:codeex}
\end{figure}

We now illustrate the advantages of our approach using a simple vector scaling example shown in Figure~\ref{fig:codeex}.
The user expresses the computation by writing a high-level expression using our \pat{map} algorithmic pattern as shown in Figure~\ref{fig:codeex:map}.
This coding style is similar to functional and dataflow programming.

Our technique first rewrites the user provided high-level expression into something closer to the OpenCL programming model.
This is achieved by applying the rewrite rules presented later in Section~\ref{rules}.
Figure~\ref{fig:codeex:impl} shows one possible derivation of the original high-level expression where the $\circ$ operator represents function composition, \ie $f \circ g(x) = f(g(x))$.
Starting from the last line, we first split the input into chunks of 1024 elements.
Each chunk is mapped onto a group of threads, called \emph{workgroup}, with the \pat{map-workgroup} low-level pattern (line~2).
Within a workgroup (lines~3--5), we vectorize the elements (line~5), each mapped to a local thread inside a workgroup via the \pat{map-local} low-level pattern (line~3).
Each local thread now processes 4 elements, enclosed in a vector type.
Finally, the \pat{vectorize-4} pattern (line~4) implies that the user defined function \code{mul3} is vectorized.
The exact meaning of our patterns will be given later in Section~\ref{pattern}.

The last step consists of traversing the low-level expression and generating OpenCL code for each low-level pattern encountered (Figure~\ref{fig:codeex:ocl}).
The two map patterns generate the for loops (line~3--4 and~7--8) that iterate over the input array assigning work to the workgroups and local threads.
The information of how many chunks each workgroup and thread processes comes from the corresponding \pat{split}.
In line~11 the vectorized version of the user defined \code{mul3} function (defined in line~1) is finally applied to the input array.

To summarize, our approach is able to generate OpenCL code starting from a high-level representation of a program.
This is achieved by systematically lowering the high-level expression into a low-level form suitable for code generation.
The next two sections present our high-level and low-level patterns, the code generation mechanism and the rewrite rules in more details.

\section{Patterns Design and Implementation}
\label{pattern}

\captionsetup[table]{margin=1.75em}
\begin{table*}[t]
\centering
\rowcolors{2}{oddcolor}{evencolor}
\begin{tabular}{lll}
\toprule
\tabhead{Pattern} & \tabhead{Type} & \tabhead{Description}\\
\midrule
 \pat{map(f)}         & $T[n] \rightarrow U[n], f: T \rightarrow U$            & Apply function $f$ to every element of the input array.\\
 \pat{reduce(f, z)}   & $T[\ ] \rightarrow T[1], f: (T,T) \rightarrow T, z : T$& Apply the reduction function $f$ with initial value z to the input array.\\
 \pat{zip(a,b)}       & $a:T[n], b:U[n] \rightarrow \langle T,U \rangle [n]$   & Zip two arrays into an array of pairs.\\
 \pat{split}$^n$      & $T[m][\ ]^* \rightarrow T[m/n][n][\ ]^*$               & Splits the outer most dimension of an array in chunks of size n.\\
 \pat{join}           & $T[m][n][\ ]^* \rightarrow T[m*n][\ ]^*$               & Joins the two outer most dimensions of an array.\\
 \pat{iterate$^n$(f)} & $T[\ ] \rightarrow T[\ ], f: T[\ ] \rightarrow T[\ ]$  & Iterate the function $f$ over the input $n$ times.\\
 \pat{reorder}        & $T[n] \rightarrow T[n]$                                & Reorder the element of the input array.\\
\bottomrule
\end{tabular}
\caption{High-level algorithmic patterns used by the programmer. $T \rightarrow U$ means the function input type is $T$ and output type $U$. We write $T[n]$ for an array of type $T$ with size $n$ and $[\ ]^*$ denotes an arbitrary number of dimensions in an array.\vspace{-1em}}
\label{tab:hlskel}
\end{table*}
\begin{table*}[t]
\centering
\rowcolors{2}{oddcolor}{evencolor}
\setlength{\tabcolsep}{2pt}
\begin{tabular}{lll}
\toprule
\tabhead{Pattern} & \tabhead{Type} & \tabhead{Description}\\
\midrule
 \pat{map-workgroup(f)}    & identical to \pat{map(f)}                     & Each \textbf{workgroup} applies function $f$ on a different element of the input array.\\
 \pat{map-local(f)}        & identical to \pat{map(f)}                     & Each \textbf{local thread} applies function $f$ on a different element of the input array.\\
 \pat{map-global(f)}        & identical to \pat{map(f)}                     & Each \textbf{global thread} applies function $f$ on a different element of the input array.\\
 \pat{map-seq(f)}          & identical to \pat{map(f)}                     & Apply function $f$ to every element of the input array \textbf{sequentially}.\\
 \pat{reduce-seq(f,z)}     & identical to \pat{reduce(f,z)}                & Apply reduction function $f$ with initial value z to the input \textbf{sequentially}.\\  
 \pat{reorder-stride}$^s$  & identical to \pat{reorder}                    & Access input array with a stride $s$ to maintain \textbf{memory coalescing}.\\
 \pat{toLocal(f)}          & $T[\ ] \rightarrow U[\ ], f: T[\ ] \rightarrow U[\ ]$  & Change the storage location for the results of function $f$ to \textbf{local memory}.\\
 \pat{toGlobal(f)}         & $T[\ ] \rightarrow U[\ ], f: T[\ ] \rightarrow U[\ ]$  & Change the storage location for the results of function $f$ to \textbf{global memory}.\\
 \pat{asVector}$^n$        & $T[\ ]^*[m] \rightarrow Tn[\ ]^*[m/n]$        & Turns the elements of an array into \textbf{vector type}.\\
 \pat{asScalar}            & $Tn[\ ]^*[m] \rightarrow T[\ ]^*[m*n]$        & Turns the elements of an array into \textbf{scalar type}.\\
 \pat{vect}$^n$\pat{(f)}   & $T[n] \rightarrow U[n], f: T \rightarrow U$   & \textbf{Vectorize} the function $f$ by a factor $n$.\\
\bottomrule
\end{tabular}
\caption{Low-level OpenCL patterns used for code generation. The hardware paradigm used is highlighted in bold in the description.}
\label{tab:llskel}
\end{table*}

One of the key ideas of this paper is to expose algorithmic choices and hardware-specific program optimizations as patterns that can be systematically derived using a rule rewriting system (discussed later in in Section~\ref{rules}).
The high-level algorithmic patterns are designed to be used by the programmer directly.
The low-level hardware patterns represent hardware specific concepts expressed by a low-level programming model such as OpenCL, the target chosen for this paper.
Following the same approach, a different set of low-level hardware patterns could be designed to target other low-level programming models such as Pthreads or MPI.

This section discusses the design of our patterns and how we generate OpenCL code for them.
We define our patterns as functions which are implicitly applied to exactly one input array and produces one output array.
To simplify our implementation we decided to encode all types as arrays with primitives represented with arrays of length 1.
The only exceptions are the user-provided functions such as the \code{mul3} function in Figure~\ref{fig:codeex:map} that operates on a primitive type.

\subsection{Algorithmic Patterns}

Table~\ref{tab:hlskel} presents our high-level algorithmic patterns.
These patterns are not tied to any specific hardware feature and are used to define the program at the algorithmic level by the programmer.


\paragraph{Map}
The \pat{map} pattern is well known in functional programming and applies a given function $f$ to all elements of its input array.

\paragraph{Reduce}
The \pat{reduce} pattern (a.k.a. fold or accumulate) uses a given binary function $f$ to combine all elements of the input array.
We require the function $f$ to be associative and commutative which allows for an efficient parallel implementation.

\paragraph{Zip and Split/Join}
These patterns transform the shape of the data and we store this information, \ie number of dimensions and size of each dimension, in the type system.
The \pat{zip} pattern fuses two arrays into an array of pairs.
The \pat{split} pattern, which is most often combined with a \pat{join}, partitions an array into chunks of specific size resulting in an extra dimension.
The corresponding \pat{join} pattern does the opposite; it reassembles arrays of arrays by merging dimensions.
These two patterns used together are similar to the split-join concept from data flow languages such as StreamIt~\cite{thies02streamit}.
 
\paragraph{Iterate}
The \pat{iterate} pattern corresponds to the mathematical definition of iteratively applying a function.
It is defined as: {$f^0 = id$} and {$f^{n+1} = f^n \circ f$}.
In terms of implementation, our code generator emits a for-loop to perform the iteration, and two pointers for input and output.
After each iteration, we swap the pointers, so that the output of the last iteration becomes the input for the next one.

\paragraph{Reorder}

The \pat{reorder} pattern is used to specify that the ordering of the elements of an array does not matter.
This allows our system to reorder arbitrarily the elements of an array and might enable optimizations, as we will see later.


\subsection{OpenCL-specific Patterns}

It is well known, that programming parallel hardware such as manycore CPUs and GPUs is quite complex.
In order to achieve the highest performance, programmers often use a set of rules of thumb to drive the optimization of their application for the specific devices.
In fact each hardware vendor provides optimization guides~\cite{nvidia11guide,amd12guide} that extensively cover hardware particularities and how to optimize code for them.

In this paper we focus on the OpenCL programming model, which is a popular low-level programming model used to program manycore CPUs and GPUs.
Programming these devices consists of writing a compute \emph{kernel} in OpenCL C that executes on the device and writing the host code that orchestrates data movement, allocates memory and manages the execution on the device.

We encode the OpenCL programming model by formalizing hardware paradigms expressed as patterns.
Table~\ref{tab:llskel} gives an overview of the OpenCL-specific patterns we have identified.

\paragraph{Parallel Maps}

The different \pat{map} patterns represent possible ways of mapping computations to the hardware and exploit parallelism in OpenCL.
The \pat{map-workgroup(f)} pattern assigns work to a group of threads, called \emph{workgroup} in OpenCL, by letting every workgroup apply the function $f$ on a different element of the input array.
Similarly, the \pat{map-local(f)} pattern assigns work to a local thread inside a workgroup.
As workgroups are optional in OpenCL the \pat{map-global(f)} pattern assigns work to a global thread, \ie a thread not organized in a workgroup.
This allows us to map computations in different ways to the thread hierarchy of OpenCL.

The code generation for all these map patterns is similar, we describe it using \pat{map-workgroup(f)} as an example.
A loop is generated, where the iteration variable is determined by the \emph{workgroup-id}, which is provided by OpenCL.
Inside of the loop, a pointer is generated to partition the input array, so that every workgroup processes a different chunk of data.
We use this pointer as the input for the function $f$ being bound to the map.
Similarly, we generate and use a pointer for the output.
After emitting the code for the loop, we continue with the body of the loop by generating the code for the function $f$.
When the generation of the body is finished, an appropriate synchronization mechanism for the given map pattern is added.
For instance after a \pat{map-local} we add a barrier synchronization to synchronize the threads inside of the workgroup.


\paragraph{Sequential Map and Reduce}
The \pat{map-seq(f)} and \pat{reduce-seq(f,~z)} patterns perform a sequential map and reduction, respectively, within a single thread.
In both cases the generated code consists of a simple for loop iterating over the array and calling the function $f$.
In case of the reduction an accumulation variable is initialized with $z$.
The variable is then passed to the function $f$ in each iteration and the computed result is stored in the accumulation variable and finally written to the output of the reduction.

\paragraph{Reorder-stride}
Using this pattern the elements of an array are reordered with a stride $s$.
In effect, this generates an access to an array such that $out[i] = in[i / n + s \cdot (i \bmod{n})]$, where $n\cdot s$ is the size of the array.
This hardware pattern ensures that after splitting the workload, consecutive threads access consecutive memory elements (known as a \emph{coalesce memory access}) which is beneficial on modern GPUs as it maximizes the memory bandwidth.

Our implementation of this pattern does not produce code directly, but generates instead an index function, which is used when accessing the array the next time.
While not discussed here, our design allows us to support user-defined index functions as well.

\paragraph{Local/Global}
The \pat{toLocal(f)} and \pat{toGlobal(f)} patterns are used to determine where the result of function $f$ should be stored.
OpenCL defines two distinct address spaces: global and local.
Global memory is the commonly used large but slow memory.
On GPUs, the small local memory has a high bandwidth with low latency and is used to store frequently accessed data.
With these two patterns, we can in effect exploit the memory hierarchy defined in OpenCL.
These patterns act similarly to a typecast and are in fact implemented as such so that no code is emitted directly.

In our design, every function reads its input and writes its output using pointers provided by the callee function.
As a result, we can simply force a store to local memory by wrapping any function with our \pat{toLocal} pattern.
In the code generator, this will simply change the output pointer of function $f$ to an area in local memory.

\paragraph{Vectorize and asVector/asScalar}
The OpenCL programming model supports vectorization with special data types such as \code{int4} where any operations on this type will be executed in the hardware vector units.
In the absence of vector units in the hardware, the OpenCL compiler scalarizes the code automatically.

The \pat{asVector} and \pat{asScalar} patterns change the data type into vector elements and scalar elements respectively.
For instance, in OpenCL an array of \code{int} is transformed into an array of \code{int4} as seen in the motivation example (Figure~\ref{fig:codeex}).
The \pat{vect$^n$(f)} pattern vectorizes a function by simply converting all the operations in $f$ that apply to vector types into vectorized operations. 
Our current implementation can only vectorize functions containing simple arithmetic operations such as $+$ or $-$.
In case of more complex functions, we rely on external tools~\cite{garrenberg11vect} for vectorizing the operations. 



\section{Rewrite Rules}
\label{rules}

This section introduces our set of rewrite rules that transform high-level expressions written using our algorithmic patterns into semantically equivalent expressions.
One goal of our approach is to keep each rule as simple as possible and only express one fundamental concept at a time.
For instance the vectorization rule, as we will see, is the only place where we express the vectorization concept.
This is different from most prior approaches that would produce a special vectorized version of different algorithmic patterns such as map or reduce.
The superiority of our approach lies in the power of composition;
many rules can be applied successively to produce expressions that compose hardware concepts or optimizations and that are provably correct by construction.

Similarly to our patterns, we distinguish between algorithmic and lowering rules.
Algorithmic rules produces derivations that represent the different algorithmic choices and are shown in Figure~\ref{fig:algo}.
Figure~\ref{fig:low} shows our OpenCL-specific rules which map expressions to OpenCL patterns.
Once the expression is in its lowest form, it is possible to produce OpenCL code for each single pattern easily with our code generator as described in the previous section.



\subsection{Algorithmic Rules}


 
\paragraph{Iterate decomposition}
The rule in Figure~\ref{fig:algo:iterate} expresses the fact an iteration can be decomposed into several iterations.

\paragraph{Reorder commutativity}
Figure~\ref{fig:algo:reorder} shows a rule stating that if the data can be reordered arbitrarily it does not matter if we apply a function $f$ to each element before or after the reordering.

\paragraph{Split-join}
The split-join rule in Figure~\ref{fig:algo:splitjoin} partitions a map into two maps.
This allows us to nest map patterns in each other and, thus, \emph{map} the computation to the thread hierarchy of the OpenCL programming model such as \pat{map-workgroup(map-local(f))} as seen in our motivation example (Figure~\ref{fig:codeex}).

\paragraph{Reduction}
The reduction (and associated partial reduction) in Figure~\ref{fig:algo:red} is currently our most complex rule but also the most powerful one.
It expresses the reduction function as a composition of other primitive functions, which is a fundamental aspect of our work.
From the algorithmic point of view we first define a partial reduction pattern \pat{part-red}.
This partial reduction reduces an array of $n$ elements to an array of $m$ elements where $1 \leq m < n$.
The reduction can be derived in a partial reduction combined with a full reduction which ensures we end up with one unique element.

\paragraph{Partial Reduction}
The first possible derivation for partial reduction, in Figure~\ref{fig:algo:red}, leads to the full reduction which means $m=1$.
The next possible derivation expresses the fact that it is possible to reorder the elements to be reduced, expressing the commutativity property of our definition of reduction.
The third derivation is actually the only place where parallelism is expressed in the definition of our reduction pattern.
This rule expressed the fact that it is valid to partition the input elements first and then reduce them independently.
Finally, the last possible derivation expresses the notion that it is possible to perform a partial reduction with an iterative process by repetitively applying the same partial reduction function.
This concept is very important when considering how the reduction function is typically implemented on a GPU (iteratively reducing within a workgroup using the local memory).

\paragraph{Simplification Rules}
Figure~\ref{fig:algo:simpl} shows our simplification rules.
They express the fact that consecutive \pat{split}-\pat{join} pairs and \pat{asVector}-\pat{asScalar} pairs are equivalent to the identity function \emph{id}.

\paragraph{Fusion Rules}
Finally, our fusion rules are shown in Figure~\ref{fig:algo:fusion}.
The first rule fuses the functions applied by two consecutive maps.
The second rule fuses the map-reduce pattern by creating a lambda function that is the results of merging function $f$ and $g$ from the original reduction and map respectively.
This rule only applies to the sequential version since this is the only implementation not requiring the associativity property required by the more generic $\pat{reduce}$ pattern.
When generating code, these rules in effect allow us to fuse the implementation of the different functions and avoid having to store temporary results.
More generic rules for fusion have been studies by the functional programming community~\cite{coutts07streamfusion,jones01playing}.
However, as we currently focus on a restricted set of patterns our simpler fusion rules have, so far, proven to be sufficient.

\newlength{\ruleSpace}
\setlength{\ruleSpace}{-.5em}
\begin{figure}[t]
\centering

\begin{subfigure}[b]{1\linewidth}
\begin{mdframed}
$$
\begin{array}{lrl}
  \pat{\textbf{iterate}}^{m+n}\pat{(f)} & \rightarrow & \pat{\textbf{iterate}}^m\pat{(f)} \circ \pat{\textbf{iterate}}^n\pat{(f)}\\
  \end{array}
$$
\end{mdframed}
  \vspace{-1em}
  \caption{Iterate decomposition}
  \label{fig:algo:iterate}
\end{subfigure}

\vspace{\ruleSpace}
\begin{subfigure}[b]{1\linewidth}
\begin{mdframed}
$$
\begin{array}{lrl}
  \pat{map(f)} \circ \pat{reorder} & \rightarrow & \pat{reorder} \circ \pat{map(f)}\\
  \pat{reorder} \circ \pat{map(f)} & \rightarrow & \pat{map(f)} \circ \pat{reorder}\\  
\end{array}
$$
\end{mdframed}
  \vspace{-1em}
  \caption{Reorder commutativity}
  \label{fig:algo:reorder}
\end{subfigure}

\vspace{\ruleSpace}
\begin{subfigure}[b]{1\linewidth}
\begin{mdframed}
$$
\begin{array}{lrl}
  \pat{map(f)} & \rightarrow & \pat{\textbf{join}} \circ \pat{map(map(f))}\circ \pat{\textbf{split}}^n
\end{array}
$$
\end{mdframed}
  \vspace{-1em}
  \caption{Split-join}
  \label{fig:algo:splitjoin}
\end{subfigure}

\vspace{\ruleSpace}
\begin{subfigure}[b]{1\linewidth}
\begin{mdframed}
$$
\begin{array}{lrl}
\pat{reduce(f,z)} & \rightarrow & \pat{reduce(f,z)} \circ \pat{part-red(f,z)}\\[.5em]
\pat{part-red(f,z)} & \rightarrow & \pat{reduce(f,z)}\\                                                      
  &| & \pat{part-red(f,z)} \circ \pat{reorder}\\    
  &| & \pat{\textbf{join}} \circ \pat{map(part-red(f,z))} \circ \pat{\textbf{split}}^n\\
  &| & \pat{\textbf{iterate}}^{n}\pat{(part-red(f,z))}\\
\end{array}
%
$$
\end{mdframed}
  \vspace{-1em}
  \caption{Reduction}
  \label{fig:algo:red}
\end{subfigure}

\vspace{\ruleSpace}
\begin{subfigure}[b]{1\linewidth}
\begin{mdframed}
$$
\begin{array}{lrl}
\pat{\textbf{split}}^n \circ \pat{\textbf{join}}^n \hspace{1em} |\hspace{1em} \pat{\textbf{join}}^n \circ \pat{\textbf{split}}^n & \rightarrow & \pat{id}\\
\pat{\textbf{asVector}}^n \circ \pat{\textbf{asScalar}}^n \hspace{1em} |\hspace{1em} \pat{\textbf{asScalar}}^n \circ \pat{\textbf{asVector}}^n       & \rightarrow & \pat{id}\\
\end{array}
$$
\end{mdframed}
  \vspace{-1em}
  \caption{Simplification rules}
   \label{fig:algo:simpl}
\end{subfigure}

\vspace{\ruleSpace}
\begin{subfigure}[b]{1\linewidth}
\begin{mdframed}
$$
\begin{array}{lll}
\pat{map(f)} \circ \pat{map(g)}                                & \rightarrow & map(f \circ g)\\
\pat{\textbf{reduce-seq}(f,z)} \circ \pat{\textbf{map-seq}(g)} & \rightarrow & \\
{\hspace{2em}} \pat{\textbf{reduce-seq}}(\lambda\ acc,x: f(acc,g(x)), z)\\  
\end{array}
$$
\end{mdframed}
  \vspace{-1em}
  \caption{Fusion rules}
   \label{fig:algo:fusion}
\end{subfigure}
\vspace{-2em}
\caption{Algorithmic rules. Bold patterns are known to the code generator.}
\label{fig:algo}
\end{figure}

\begin{figure}[t]
\centering


\begin{subfigure}[b]{1\linewidth}
\begin{mdframed}
$$
\begin{array}{lllll}
\pat{map(f)} & \rightarrow & \pat{\textbf{map-workgroup}(f)} & | & \pat{\textbf{map-local}(f)}\\
 & | & \pat{\textbf{map-global}(f)} & | & \pat{\textbf{map-seq}(f)}\\          
\end{array}
$$
\end{mdframed}
  \vspace{-1em}
  \caption{Map}
  \label{fig:low:map}
\end{subfigure}

\vspace{\ruleSpace}
\begin{subfigure}[b]{1\linewidth}
\begin{mdframed}
$$
\begin{array}{lll}
  \pat{reduce(f,z)} & \rightarrow & \pat{\textbf{reduce-seq}(f,z)}
\end{array}
$$
\end{mdframed}
  \vspace{-1em}
  \caption{Reduction}
  \label{fig:low:red}
\end{subfigure}

\vspace{\ruleSpace}
\begin{subfigure}[b]{1\linewidth}
\begin{mdframed}
$$
\begin{array}{lllll}
  \pat{reorder}  & \rightarrow & \pat{\textbf{reorder-stride}}^s & | & \pat{id}
\end{array}
$$
\end{mdframed}
  \vspace{-1em}
  \caption{Stride accesses or normal accesses}
  \label{fig:low:stride}
\end{subfigure}

\vspace{\ruleSpace}
\begin{subfigure}[b]{1\linewidth}
\begin{mdframed}
$$
\begin{array}{lrl}
  \pat{\textbf{map-local}(f)} & \rightarrow & \pat{\textbf{toGlobal}(\textbf{map-local}(f))}\\  
  \pat{\textbf{map-local}(f)} & \rightarrow & \pat{\textbf{toLocal}(\textbf{map-local}(f))}\\  
\end{array}
$$
\end{mdframed}
  \vspace{-1em}
  \caption{Local/Global memory}
  \label{fig:low:mem}
\end{subfigure}

\vspace{\ruleSpace}
\begin{subfigure}[b]{1\linewidth}
\begin{mdframed}
$$
\begin{array}{lrll}
\pat{map(f)} & \rightarrow & \pat{\textbf{asScalar}} \circ \pat{map(\textbf{vect}}^n\pat{(f))} \circ \pat{\textbf{asVector}}^n
\end{array}
$$
\end{mdframed}
  \vspace{-1em}
  \caption{Vectorization}
   \label{fig:algo:vect}
\end{subfigure}
\vspace{-2em}
\caption{OpenCL-specific rules. Bold patterns are known to the code generator.}
\label{fig:low}
\end{figure}

\subsection{OpenCL-Specific Rules}

Figure~\ref{fig:low} shows our OpenCL-specific rules that are used to apply OpenCL optimizations and to lower high-level concepts down to OpenCL-specific ones.
Patterns that are known to the code generator are shown in bold in both Figure~\ref{fig:algo} and~\ref{fig:low}.


\paragraph{Maps}
The rule in Figure~\ref{fig:low:map} is used to produce OpenCL-specific map implementations that match the thread hierarchy of the OpenCL programming model.
Our implementation maintains context information to ensure the thread hierarchy is respected.
For instance, it is only legal to nest a \pat{map-local} inside a \pat{map-workgroup}.



\paragraph{Reduction}
There is only one lowering rule for reduction (Figure~\ref{fig:low:red}), which expresses the fact that the only OpenCL implementation known to the code generator is a sequential reduction.
Possible parallel implementations of the reduction pattern are defined at a higher level by composition of other algorithmic patterns.
To the best of our knowledge, all other existing high performance compilers treat the reduction directly as an irreducible primitive operation.
The power of our approach is that the code generator implementation only needs to know about the simple sequential reduction.
As a result, it is possible to explore different implementation for the reduction by simply applying different rules.

\paragraph{Reorder}
Figure~\ref{fig:low:stride} presents the rule that reorders elements of an array.
In our current implementation, we support two types of reordering:
no reordering, represented by the \pat{id} identify function, and \pat{reorder-stride} which reorders elements with a certain stride $s$.
As described earlier, the major use case for the stride reorder is to enable coalesced memory accesses.

\paragraph{Local/Global}
Figure~\ref{fig:low:mem} shows two rules that enable GPU local memory usage.
They express the fact that the result of a \pat{map-local} can always be stored in local memory or back in global memory.
This holds since a \pat{map-local} always exists within a \pat{map-workgroup} for which the local memory is defined.
These rules allow us to determine how the data is mapped to the GPU memory hierarchy.

\paragraph{Vectorization}
Finally, Figure~\ref{fig:algo:vect} shows the vectorization rule.
Vectorization is achieved by using the \pat{asVector} and corresponding \pat{asScalar} which changes the element type of an array and adjust the length accordingly.
This rule is only allowed to be applied once to a given \pat{map(f)} pattern.
This constrain can easily be checked by looking at the function's type; if it is a vector type, the rule cannot be applied.
Another set of rules, not shown here for space reason, are used to propagate the $vect^n$ function recursively within $f$.

%
%

\subsection{Summary}

The power of our approach lies in the composition of our rules that produce complex low-level expressions from simple high-level expressions.
Looking back at our motivation example in Figure~\ref{fig:codeex}, we see how a simple algorithmic pattern such as \pat{map} can effectively be derived into a low-level expression by applying the rules.
This expression matches various hardware concepts expressible with the OpenCL programming model such as mapping computation and data to the GPU thread and memory hierarchy and vectorization.
Each single rule encodes a simple, easy to understand, provable fact.
By composition of the rules we systematically derive low-level expressions which are semantically equivalent to the high-level expressions by construction.
This results in a powerful mechanism to safely explore the space of possible implementations.


\section{Benchmarks}
\label{benchmarks}

We now discuss how applications from linear algebra, mathematical finance and physics can be represented as expressions composed of our high-level algorithmic patterns.
We use the following conventions to simplify the syntax:
non-capitalized letters (\eg \code{x}) denote scalar variables,
letters with an arrow on top (\eg $\vec{x}$) denote 1D vectors, and
capitalized letters (\eg $A$) denote 2D matrices.

\subsection{Linear Algebra Kernels}

\begin{figure}[t]
\begin{lstlisting}[mathescape,numbers=left,xleftmargin=2.5em]
def add(x, y)  = x + y
def mult(x, y) = x * y
def abs(x)     = if (x < 0) -x else x

def scal(a, $\vec{x}$) = map(mult(a), $\vec{x}$)
def asum($\vec{x}$)     = reduce(add, 0)$\,\circ\,$map(abs, $\vec{x}$)
def dot($\vec{x}$, $\vec{y}$)    = reduce(add, 0)$\,\circ\,$map(mult)$\,\circ\,$zip($\vec{x}$,$\,\vec{y}$)
def gemv($A$, $\vec{x}$, $\vec{y}$, a, b) =
   $\vec{z}$ = map(scal(a) $\circ$ dot($\vec{x}$), $A$)
   map(add) $\circ$ zip( $\vec{z}$, scal(b, $\vec{y}$) )
\end{lstlisting}
\caption{Linear algebra kernels from the BLAS library expressed using our high-level algorithmic patterns.}
\label{fig:linearAlgebra}
\end{figure}

We choose linear algebra kernels as our first set of benchmarks, because they are well known, easy to understand, and used as building blocks in many other applications.
Figure~\ref{fig:linearAlgebra} shows how we express vector scaling (line~5), sum of absolute values (line~6), dot product of two vectors (line~7) and matrix vector multiplication (line 8--10) using our high-level patterns.
While the first three benchmarks perform computations on vectors, matrix vector multiplication was chosen to illustrate a computation using a 2D data structures.

For scaling (line~5), the \pat{map} pattern applies a function to each element which multiplies it with a constant.
This function is expressed by partially applying the \code{mult} function, \ie binding \code{a} to the first argument of \code{mult}.
The sum of absolute values (line~6) and the dot product (line~7) applications both produce scalar results by performing a summation, which we express using the \pat{reduce} pattern combined with the addition.
For dot product, a pair-wise multiplication of its two input vectors is performed before applying the reduction.
This is expressed using the \pat{zip} and \pat{map} patterns.

Line~8--10 shows the implementation of matrix vector multiplication as defined by the BLAS library: $\vec{y} = \alpha A \vec{x} + \beta \vec{y}$.
To multiply matrix $A$ with vector $\vec{x}$, the \pat{map} pattern maps the computation of the dot-product with the input vector $\vec{x}$ to each row of the matrix $A$ (line~9).
Notice how we are reusing the high-level expressions for dot-product and scaling as building blocks for the more complex matrix-vector multiplication.
This shows the power of our system: expressions describing algorithmic concepts can be reused, without committing to a particular low-level implementation;
The dot-product from gemv (line~9) might be implemented in a totally different way from the stand-alone dot-product kernel (line~7).

\subsection{Mathematical Finance Application}

The BlackScholes application uses a Monte-Carlo method for option pricing and computes for each stock price \code{s} a pair of call and put options \code{\{c, p\}}.
Figure~\ref{fig:blackScholes} shows the BlackScholes implementation, where the function defined in line~1 computes the call and put option for a single stock price \code{s}.
Two intermediate results \code{d1} and \code{d2} are computed and used to compute the call and put options which are returned as a single pair.
The \code{compD1}, \code{compD2}, \code{compCall} and \code{compPut} functions are not shown here since they only contain purely sequential code implementing the BlackScholes model.
This \code{BSComputation} function is applied to all stock prices, stored in a vector $\vec{s}$, using the \pat{map} pattern in line~4.

\begin{figure}[t]
\begin{lstlisting}[mathescape,numbers=left,xleftmargin=2em]
def BSComputation(s) =
  d1 = compD1(s); d2 = compD2(d1,s)
  return { compCall(d1,d2,s), compPut(d1,d2,s) }
def blackScholes($\vec{s}$) = map(BSComputation, $\vec{s}$)
\end{lstlisting}
\caption{BlackScholes mathematical finance application expressed using our high-level algorithmic patterns.}
\label{fig:blackScholes}
\end{figure}

\subsection{Physics Application}

\begin{figure}[t]
\begin{lstlisting}[mathescape,numbers=left,xleftmargin=2em]
def updateF(f, nId, p, $\vec{p}$, t) =
  n = $\vec{p}$[nId];   d = calculateDistrance(p, n)
  if (d < t) f += calculateForce(d)
  return f
def md($\vec{p}$, $N$, t) = map(
  $\lambda$ p,$\,\vec{n}$: reduce($\lambda$ f,$\,$nId:$\,$updateF(f,nId,p,$\vec{p}$,t), 0, $\vec{n}$)
 ) $\circ$ zip($\vec{p}$, $N$)
\end{lstlisting}
\caption{Molecular dynamics physics application expressed using our high-level algorithmic patterns.}
\label{fig:md}
\end{figure}

Another application we consider is the the molecular dynamics (MD) application from the SHOC~\cite{danalis10shoc} benchmark suite.
It calculates the sum of all forces acting on a particle from its neighbors.
Figure~\ref{fig:md} shows the implementation using our high-level patterns.

The function \code{updateF} is defined in line~1 and updates the force \code{f} of particle \code{p} by computing and adding the local force between a single particle and one of its neighbors.
\code{updateF} takes an index of a neighbor \code{nId}, the vector storing all particles $\vec{p}$, and a threshold \code{t} as additional parameters.
Using \code{nId} and $\vec{p}$ the neighboring particle is accessed in line~2 and the distance between the neighboring particle and the particle \code{p} ist computed.
If the distance is below the given threshold \code{t} the local force between the two particles is calculated based on the distance and added to the overall force \code{f} (line~3) which is finally returned in line~4.
Otherwise the particle is ignored in the summation.

\newcommand{\Reduce}{\text{\textit{reduce}}\xspace}
\newcommand{\PartRed}{\text{\textit{part-red}}\xspace}
\newcommand{\RedSeq}{\text{\textit{reduce-seq}}\xspace}
\newcommand{\Map}{\text{\textit{map}}\xspace}
\newcommand{\MapSeq}{\text{\textit{map-seq}}\xspace}
\newcommand{\MyJoin}{\text{\textit{join}}\xspace}
\newcommand{\MySplit}[1]{\text{\textit{split}}^{#1}\xspace}

\begin{figure*}[t]
\begin{align}
  \text{\textit{asum}}(\vec{x})
  & =\hspace{.2em} \Reduce(+, 0) \circ \Map(abs, \vec{x})\\[-.5em]
  & \overset{\ref{fig:algo:red}}{=\hspace{.2em}}
      \Reduce(+, 0) \circ \MyJoin \circ \Map(\PartRed(+, 0)) \circ \MySplit{n} \circ \Map(abs, \vec{x})\\[-.5em]
  & \overset{\ref{fig:algo:splitjoin}}{=\hspace{.2em}}
      \Reduce(+, 0) \circ \MyJoin \circ \Map(\PartRed(+, 0)) \circ \MySplit{n} \circ \MyJoin \circ \Map(\Map(abs)) \circ \MySplit{n}(\vec{x})\\[-.5em]
  & \overset{\ref{fig:algo:simpl}}{=\hspace{.2em}}
      \Reduce(+, 0) \circ \MyJoin \circ \Map(\PartRed(+, 0)) \circ \Map(\Map(abs)) \circ \MySplit{n}(\vec{x})\\[-.5em]
  & \overset{\ref{fig:algo:fusion}}{=\hspace{.2em}}
      \Reduce(+, 0) \circ \MyJoin \circ \Map(\PartRed(+, 0) \circ \Map(abs)) \circ \MySplit{n}(\vec{x})\\[-.5em]
  & \overset{\ref{fig:low:map}}{=\hspace{.2em}}
      \Reduce(+, 0) \circ \MyJoin \circ \Map(\PartRed(+, 0) \circ \MapSeq(abs)) \circ \MySplit{n}(\vec{x})\\[-.5em]
  & \overset{\hspace{-1.3em}\ref{fig:algo:red}\& \ref{fig:low:red}\hspace{-1.1em}}{=\hspace{.2em}}
      \Reduce(+, 0) \circ \MyJoin \circ \Map(\RedSeq(+, 0) \circ \MapSeq(abs)) \circ \MySplit{n}(\vec{x})\\[-.5em]
  & \overset{\ref{fig:algo:fusion}}{=\hspace{.2em}}
      \Reduce(+, 0) \circ \MyJoin \circ \Map(\RedSeq(\lambda\ acc, a: acc + abs(a), 0) \circ \MySplit{n}(\vec{x})
\end{align}
\vspace{-2em}
\caption{Derivation for \emph{asum}$(\vec{x})$ to a fused version.
  The numbers above the equality sign refer to the rules from Figure~\ref{fig:algo} and Figure~\ref{fig:low}.
}
\label{fig:derivation}
\end{figure*}
\begin{figure*}[t]
\vspace{-1.0em}
\captionsetup[subfigure]{justification=justified,singlelinecheck=false}

\begin{subfigure}[b]{\linewidth}
\vspace{.4em}
\begin{minipage}{.01\linewidth}
\caption{}
\label{fig:llexpr:nvidia}
\end{minipage}
\hfill
\begin{minipage}{.96\linewidth}
\begin{lstlisting}[mathescape]
def asum($\vec{x}$) = reduce-seq o join o map-workgroup(
     join o toGlobal( map-local( map-seq(id) ) ) o split-1
   o iterate-7( join o map-local( reduce-seq(plus, 0) ) o split-2 )
   o join o toLocal( map-local( reduce-seq(absAndPlus, 0) ) ) o split-2048 o reorder-stride
 ) o split-262144($\vec{x}$)
\end{lstlisting}
\end{minipage}
\end{subfigure}

\begin{subfigure}[b]{\linewidth}
\vspace{0em}
\begin{minipage}{.01\linewidth}
\caption{}
\label{fig:llexpr:amd}
\end{minipage}
\hfill
\begin{minipage}{.96\linewidth}
\begin{lstlisting}[mathescape]
def asum($\vec{x}$) = reduce-seq o join o asScalar o map-workgroup(
     join o toGlobal( map-local( map-seq(vectorize-4(id) ) ) ) o split-1
   o iterate-8( join o map-local( reduce-seq(vectorize-4(plus), vectorize-4(0)) ) o split-2 )
   o join o toLocal( map-local( reduce-seq(vectorize-4(absAndPlus), vectorize-4(0)))) o split-2 o reorder-stride 
 ) o asVector-4 o split-2048($\vec{x}$)
\end{lstlisting}
\end{minipage}
\end{subfigure}

\begin{subfigure}[b]{\linewidth}
\vspace{0em}
\begin{minipage}{.01\linewidth}
\caption{}
\label{fig:llexpr:intel}
\end{minipage}
\hfill
\begin{minipage}{.96\linewidth}
\begin{lstlisting}[mathescape]
def asum($\vec{x}$) = reduce-seq o join $\circ$ asScalar $\circ$ map-workgroup(
   join o map-local( reduce-seq(vectorize-4(absAnd+), vectorize-4(0)) ) o split-8192
 ) o asVector-4 o split-32768($\vec{x}$)
\end{lstlisting}
\end{minipage}
\end{subfigure}
\vspace{-0.5em}
\caption{Low-level expressions performing the sum of absolute values specialized for Nvidia (a), AMD (b), and Intel (c).
         These expressions are systematically derived by our system from the high-level expression $\Reduce(+, 0) \circ \Map(abs, \vec{x})$.}
\label{fig:llexpr}
\end{figure*}

For computing the force for all particles $\vec{p}$, we use the \pat{zip} pattern~(line~7) to build a vector of pairs.
Each pair combines a single particle with the indices of all of its neighboring particles.
The function which is applied to each pair by the \pat{map} pattern~(line~5) is expressed as an lambda expression~(line~6).
Computing the resulting force exerted by all the neighbors on one particle is done by applying the \pat{reduce} pattern on vector $\vec{n}$ which stores the indices of the neighboring particles.
We use the previously defined function \code{updateF} inside the reduction to compute the force each particle with index \code{nId} add to the overall force on \code{p}.
At this point we fix all but the first two arguments as the other arguments remain constant for particle \code{p}.
The usage of lambda expressions in our system allows for easy binding of additional information as arguments to functions.
This application example should give some evidence that our patterns are flexible enough to implement real world applications.


\section{Deriving Specialized Implementations}
\label{application}

This section shows how our rules can be applied to derive different implementations starting from the same high-level expression.
We illustrate this process using the \emph{asum} benchmark from the previous section as a simple example.
The computation can easily be expressed using two of our high-level algorithmic patterns, as shown in Figure~\ref{fig:derivation} (1).
The $abs$ function is applied to every element of the input vector $\vec{x}$ and then the intermediate result is summed up using the \Reduce pattern which is customized with the addition operator.

\subsection{Deriving a Fused Implementation}
\label{sec:example}

To achieve good performance it is in general beneficial to avoid storing intermediate results.
Rule~\ref{fig:algo:fusion} allows us to apply this principle and fuse two patterns into one, thus, avoiding an intermediate result.
Figure~\ref{fig:derivation} shows how we can systematically derive a fused version of the \emph{asum} application from the high-level expression written by the programmer.
We write the derivation as a sequence of equations using a slightly more mathematical notation, where the numbers above the equality sign refer to the rules applied.

To obtain expression (2) we apply the reduction rule~\ref{fig:algo:red} twice:
first to replace \pat{reduce} with \pat{reduce}~$\circ$~\pat{part-red} and then a second time to expand \pat{par-red}.
Afterwards, we expand \pat{map} to get (3), which can be simplified by removing the two corresponding \pat{join} and \pat{split} patterns.
In the step from (4) to (5) two \pat{map} patterns are fused and in the next step the nested \pat{map} is lowered into the \pat{map-seq} pattern to obtain (6).
By first transforming \pat{part-red} back into \pat{reduce} (using rule~\ref{fig:algo:red}) and then applying the lowering rule~\ref{fig:low:red} we get (7).
Finally, we apply rule~\ref{fig:algo:fusion} to fuse the \pat{map-seq} and \pat{reduce-seq} into a single \pat{reduce-seq}.
This sequence of transformations results in expression (8) which allows for a more optimal implementation since no temporary storage is required for the intermediate result.

\subsection{Deriving Device Specific Implementations}
\label{sec:deviceSpecific}

The previous section showed how an optimization can be systematically applied which is generally beneficial on every hardware platform.
However, there exist many optimizations which are highly specific to a particular hardware architecture.
For instance it is often beneficial to apply vectorization on an Intel CPU but not on an Nvidia GPU, on the contrary using local memory is usually beneficial on GPUs but not on CPUs.
Figure~\ref{fig:llexpr} shows three different implementations of the \emph{asum} benchmark which have been derived using the same systematic approach of applying rules as seen in Figure~\ref{fig:derivation}.
These implementations have been in fact inspired by hand-tuned OpenCL and CUDA kernels from the different vendors.
This demonstrates the expressive power of our OpenCL patterns.
Each implementation is optimized to take advantage of the features of a particular hardware architecture.
The integer parameters deciding how to split the data and the width of vectorization where chosen by exploring different values empirically.

The first implementations, shown in Figure~\ref{fig:llexpr:nvidia}, is optimized for an Nvidia GPU.
The input vector $\vec{x}$ is split in large chunks which are processed in parallel by different workgroups.
Inside each workgroup the \code{reorder-stride} pattern ensures fast coalesced memory accesses when loading the data from global memory.
Each local thread reduces 2048 elements and stores the intermediate result in local memory.
Afterwards, the entire workgroup performs an iterative computation to reduce the intermediate results down to a single result before this is copied back to global memory.
Figure~\ref{fig:llexpr:amd} shows the AMD optimized implementation which is similar to the previous one.
The same set of optimizations have been applied to take advantage of the local memory and to ensure coalesced memory accesses.
In addition, since the AMD GPU has vector units, the reduction has been vectorized by a width of four.

The third implementation as seen in Figure~\ref{fig:llexpr:intel}, is targeted at an Intel CPU and is very different.
It neither uses local memory nor the \code{reorder-stride} pattern.
Vectorization is applied, similar to the AMD version, but in contrast, the implementation uses different numbers for partitioning the data for workgroup and local threads;
only a single thread is active inside each workgroup
This corresponds to the fact, that there is less parallelism available on a CPU compared to GPUs.
These three different implementations derived from the same high-level expression should give some evidence of the power of our approach which is able to systematically derive highly hardware-specific implementations.

\subsection{Towards Automatic Derivation}
 
We have shown how our system can systematically transform and optimize programs at an algorithmic level and at a hardware level without performing complex compiler analysis.
While manually deriving the expression is a tedious process, our vision is for this to take place in a fully automated way following the principles introduced in our work.
Our rules and OpenCL-specific patterns pave the way to a fully automatic search strategy starting from the high-level expression. 
There has been other work in this area ranging from random search to sophisticated search strategies based on performance models or machine learning techniques~\cite{demesmay09bandit, phothilimthana13portable}.
We see this work as completely orthogonal to this paper, since our first focus is to develop the systematic foundations necessary to apply such techniques.
Nonetheless, we have implemented a prototype automatic search technique that is actually able to find expressions with similar performance to those presented in Figure~\ref{fig:llexpr}.
This suggest it is possible to automatically derive highly tuned implementations from high-level expressions but this is left for future work.





\section{Experimental Setup}
\label{sec:setup}

This section describes some implementation details of our code generator and our experimental setup used for the experiments.

\subsection{Implementation Details}

Our system is implemented in C++11, using the template system and support for lambda functions. 
When generating code for a derived expression two basic steps are performed.
First, we use the Clang/LLVM compiler library to parse the input expression and produce an abstract syntax tree for it.
Second, we traverse the tree and emit code for every function call representing one of our low-level hardware patterns.

As part of the first step, we have developed our own type system which plays a dual role.
First, it prevents the user to produce incorrect expressions.
Secondly, the type system encodes information necessary for code generation, such as memory address space and array size information, which are used to allocate memory.

The design of our code generator is straightforward since no optimization decisions are made at this stage.
We avoid performing complex analysis of the code which makes our design very different compared to traditional optimizing compilers.

\subsection{Hardware Platforms and Evaluation Methodology}

We used three hardware platforms to perform the runtime experiments: an Nvidia GeForce GTX 480 GPU, an AMD Radeon HD 7970 GPU and a dual socket Intel Xeon E5530 server, with 8 cores in total and hyper-threading enabled.
We used the latest OpenCL runtime from Nvidia (CUDA-SDK 5.5), AMD (AMD-APP 2.8.1) and Intel (XE 2013 R3 3.2.1.16712).
The GPU drivers installed on our Linux system were 310.44 for Nvidia and 13.1 for AMD.

We use the profiling APIs from OpenCL and CUDA to measure kernel execution time and the \textit{gettimeofday} function for the CPU implementation.
Following the Nvidia benchmarking methodology~\cite{Harris07reduction}, the data transfer time to and from the GPU is excluded.
We repeat each experiment 1000 times and report median runtimes.

For our linear algebra benchmarks, we have performed experiments with two input sizes.
For \bench{scal}, \bench{asum} and \bench{dot}, the small input size corresponds to a vector size of 16M elements (64MB).
The large input size uses 128M elements (512MB, the maximum OpenCL buffer size for our platforms).
For \bench{gemv}, we use an input matrix of 4096$\times$4096 elements (64MB) and a vector size of 4096 elements (16KB) for the small input size.
For the large input size, the matrix size is 8192$\times$16384 elements (512MB) and the vector size 8192 elements (32KB).
For \bench{BlackScholes}, the problem size is fixed to 4 million elements and for \bench{MD} it is 12288 particles.



\section{Results}
\label{sec:results}

We now evaluate our approach compared to reference OpenCL implementations of our benchmarks on all platforms.
Furthermore, we compare the BLAS routines against platform-specific tuned implementations.

\subsection{Comparison vs. Portable Implementation}

\begin{figure}[t]
  \includegraphics[width=\linewidth]{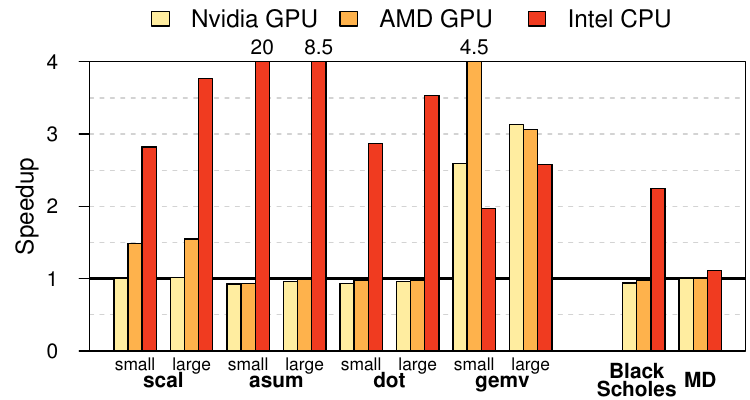} 
  \caption{Performance of our approach relative to a portable OpenCL reference implementation.
           We outperform the clBLAS implementation on most benchmarks and platforms.}
  \label{fig:clblas}
\end{figure}

We want to show how our approach performs across three platforms.
We use the BLAS OpenCL implementations written by AMD as our baseline for this evaluation since it is inherently portable across our different platforms.
Figure~\ref{fig:clblas} shows the performance of our approach relative to clBLAS for the BLAS routines. 
As can be seen, we achieve better performance than clBLAS on most platforms and benchmarks.
The speedups are the highest on the CPU, with up to 20$\times$ for the \bench{asum} benchmark with a small input size.
The reason is that clBLAS was written and tuned specifically for an AMD GPU which usually exhibit a larger number of parallel processing units.
As we saw in Section~\ref{sec:deviceSpecific}, our systematically derived expression for this benchmark is specifically tuned for the CPU by avoiding creating too much parallelism, which is what gives us such large speedup.

Figure~\ref{fig:clblas} also shows the results we obtain relative to the Nvidia SDK \bench{BlackScholes} and SHOC molecular dynamics \bench{MD} benchmark.
For \bench{BlackScholes}, we see that our approach is on par with the performance of the Nvidia implementation on both GPUs.
On the CPU, we actually achieve a 2.2$\times$ speedup due to the fact that the Nvidia implementation is tuned for GPUs while our implementation generates different code for the CPU.
For \bench{MD}, we are actually on par with the OpenCL implementation on all platforms.

\subsection{Comparison vs. Highly-tuned Implementations}

\begin{figure*}[t]
  \centering
  \begin{subfigure}[b]{0.315\linewidth}
    \includegraphics[width=\linewidth]{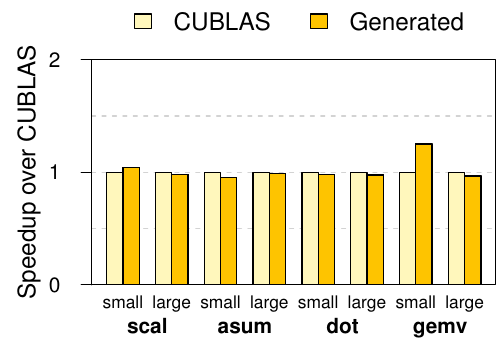}
    \caption{Nvidia GPU}
    \label{fig:results-nv}
  \end{subfigure}
  \hspace{0.015\linewidth}
  \begin{subfigure}[b]{0.315\linewidth}
    \includegraphics[width=\linewidth]{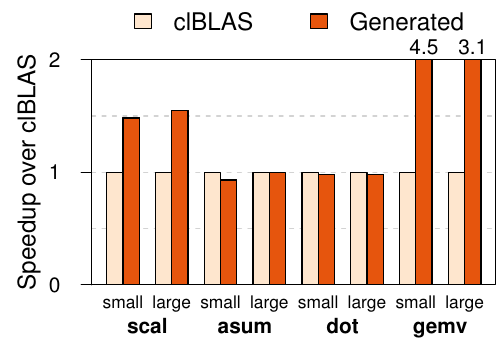} 
    \caption{AMD GPU}
    \label{fig:results-amd}
  \end{subfigure}
  \hspace{0.015\linewidth}
  \begin{subfigure}[b]{0.315\linewidth}
    \includegraphics[width=\linewidth]{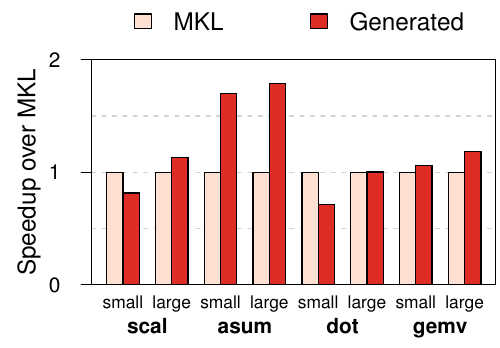} 
    \caption{Intel CPU}
    \label{fig:results-cpu}
  \end{subfigure}
  \vspace{-1.5em}
  \caption{Performance comparison of our approach relative to a highly-tuned platform-specific library; CUBLAS for Nvidia, clBLAS for AMD and MKL for the CPU.	  
           Our approach matches the performance of CUBLAS and MKL, and outperforms clBLAS on some routines.}
   \label{fig:results}  
\end{figure*}

We now compare our approach with a highly-tuned implementation for each platform.
For Nvidia, we pick the highly tuned CUBLAS CUDA-specific implementation of BLAS written by Nvidia.
For the AMD GPU, we use the same clBLAS implementation as before given that it has been written and tuned specifically for AMD GPUs.
Finally, for the CPU we use the Math Kernel Library (MKL) implementation of BLAS written by Intel which is known for its high performance.

Figure~\ref{fig:results-nv} shows that we actually match the performance of CUBLAS for \bench{scal}, \bench{asum} and \bench{dot} on the Nvidia GPU.
For \bench{gemv} we outperform CUBLAS on the small size by 20\% while we are within 5\% for the large input size.
Given that CUBLAS is a proprietary library highly tuned for Nvidia GPUs, these results should offer some confidence that our technique is able to achieve high performance.

On the AMD GPU, we are surprisingly up to 4.5$\times$ faster than the clBLAS implementation on \bench{gemv} small input size as shown in Figure~\ref{fig:results-amd}.
The reason for this is found in the way clBLAS is implemented.
For the \bench{gemv} benchmark, clBLAS performs automatic code generation using fixed templates.
In contrast to our approach they only generate one implementation since they do not explore different pattern compositions.

For the Intel CPU (Figure~\ref{fig:results-cpu}), we see that our approach beats MKL for one benchmarks and match the performance of MKL on most of the other three benchmarks.
For the small input sizes for the \bench{scal} and \bench{dot} benchmarks we are within 13\% and 30\% respectively.
For the larger input sizes we are on par with MKL for both benchmarks.
The \bench{asum} implementation in the MKL does not use thread level parallelism, where our implementation does and, thus, achieves a speedup of up to 1.78 on the larger input size.

This section has shown how our approach can generate \emph{performance portable} code that is competitive with highly-tuned platform specific implementations.





\section{Related Work}
\label{related}

\paragraph{Algorithmic Patterns}

Algorithmic patterns or skeletons~\cite{cole88skeleton} have been around for more than two decades.
Pattern-based libraries for platforms ranging from cluster systems~\cite{Rodrigues14triolet} to GPUs~\cite{Steuwer11skelcl} have been proposed with recent extension to irregular algorithms~\cite{gonzalez14irregular}.
This includes popular framework such as Map-Reduce~\cite{dean08mapreduce} from Google.
Many researchers have looked at the problem of optimizing map-reduce operations for different type of hardware.
Paraprox~\cite{Samadi14Paraprox} for instance uses automatic detection of algorithmic patterns to apply optimization at the expense of accuracy.
Compared to our approach, most prior works rely on hardware-specific implementations to achieve high performance.
Conversely, we systematically generate implementations using fine-grain OpenCL patterns combined with our rule rewriting system.

\paragraph{Functional Approaches for GPU Code Generation}

Accelerate is a functional domain specific language built within Haskell to support GPU acceleration~\cite{chakravarty11accelerating,mcdonell13optimising}.
Recently, Nvidia has presented NOVA~\cite{collins14nova}, a new functional language target at code generation for GPUs, and Copperhead~\cite{catanzaro11copperhead}, a data parallel language embedded in Python.
NOVA shares many concepts from Accelerate and Copperhead and offers familiar data parallel patterns.
HiDP~\cite{zhang13hidp} is a hierarchical data parallel language which maps computations to the OpenCL programming model similar to our approach.
All these projects rely on analysis of user code or hand-tuned versions of high-level algorithmic patterns.
In contrast, our approach uses rewrite rules and low-level hardware patterns to produce high-performance code in a portable way.

Halide~\cite{ragan-kelley13halide} is a domain specific approach targeting image processing pipelines.
It separates a programs functional algorithmic description from optimization decisions and applies autotuning to find the best optimization on different hardware platforms.
Our work is domain agnostic and takes a different approach to achieve high performance.
We systematically describe hardware paradigms as functional patterns instead of encoding specific optimizations which might not apply to future hardware generations.

\paragraph{Rewrite-rules for Optimizations}

Rewrite rules have been used very early as a way to automate the optimization process of functional programs~\cite{jones01playing}.
Recently, rewriting has been applied to HPC applications~\cite{panyala14rewriting} as well, where the user annotates imperative code providing information necessary for the rewrite process.
Similar to us, Spiral~\cite{pueschel05spiral} also uses rewrite rules to optimize signal processing programs and was more recently adapted to linear algebra~\cite{Spampinato13LGen} and other mathematical domains~\cite{Franchetti09OL}.
In contrast our rules and OpenCL hardware patterns are expressed at a much finer level, allowing for highly specialized and optimized code generation.

\paragraph{High-level Code Generation for GPUs} 

A large body of work has explored how to automatically generate high performance code for GPUs.
Dataflow programming models such as IBM's LiquidMetal~\cite{dubach12compiling} or StreamIt~\cite{thies02streamit} have been used to automatically produce GPU code with OpenCL or CUDA~\cite{hormati11sponge,huynh12scalable,udupa09software}.
Directive based approach have also been used such as OpenMP to CUDA~\cite{lee09openmp}, OpenACC to OpenCL~\cite{reyes12openaccgpu}, or hiCUDA~\cite{han11hicuda} which translates sequential C code to CUDA.
Optimized implementations for directive based reductions on GPUs has been presented~\cite{Xu14reduction} as well.
X10~\cite{tardieu14x10}, a language for high performance computing, can also be used to program GPUs~\cite{cunningham11gpu}.
However, the programming style remains low-level since the programmer has to express the same low-level operations found in CUDA or OpenCL.
Recently, researchers have looked at generating efficient GPU code for loops using the polyhedral framework~\cite{verdoolaege13polyhedral, Grosser14tiling}.
Delite~\cite{brown11heterogeneous,chafi11domain}, a system that enables the creation of domain-specific languages, can also target multicore CPUs or GPUs.
Unfortunately, all these approaches do not provide full performance portability since the mapping of the application assumes a fixed platform and the optimizations and implementations are targeted at a specific device.

Finally, Petabricks~\cite{ansel09petabricks} takes a different approach by letting the programmer specify different algorithms implementations.
The compiler and runtime then choose the most suitable one based on an adaptive mechanism and can produce OpenCL code~\cite{phothilimthana13portable}.
Compared to our work, they generate optimized code by relying on static analysis.
Our code generator does not make any decisions nor perform any analysis since the optimization process happens at a higher level within our rewrite rules.


\section{Conclusion}
\label{conc}

In this paper, we have presented a novel approach based on rewrite rules to represent algorithmic principles as well as low-level hardware-specific optimization.
We have shown how these rules can be systematically applied to transform a high-level expression into a device-specific implementation.
This results in a clear separation of concern between high-level algorithmic concepts and low-level hardware optimizations which pave the way for fully automated high performance code generation.

To demonstrate the power of our approach in practice, we have developed OpenCL-specific rules and patterns together with an OpenCL code generator.
The design of the code generator is straight-forward given that all optimizations decisions are made with the rules and no complicated analysis passes are needed.
We achieve performance on par with highly tuned platform-specific BLAS libraries on three different devices; AMD GPU, Nvidia GPU and Intel CPU.
For benchmarks such as matrix vector multiplication we even reach a speedup of up to 4.5.
We also show that our technique achieves portable performance for more complex applications such as the BlackScholes benchmark or for molecular dynamics simulation.






\balance
\bibliographystyle{abbrvnat-short-conf} 
\bibliography{main}

\end{document}